\documentclass[final,3p,times]{elsarticle}

\usepackage[colorlinks=true,linkcolor=blue]{hyperref}

\journal{{\normalfont Accepted manuscript, Engineering Fracture Mechanics, \urlstyle{same} \url{https://doi.org/10.1016/j.engfracmech.2017.02.026}}}

\usepackage{siunitx}
\usepackage{amsfonts,amsmath,amssymb,stmaryrd}
\usepackage{subcaption}
\usepackage{graphicx}

\newcommand{\bs}[1]{\boldsymbol{#1}}
\newcommand{\bt}[1]{\mathbf{#1}}
\newcommand{\tx}[1]{\text{#1}}
\newcommand{\txsc}[1]{\textsc{#1}}

\newcommand{\eps}[0]{\varepsilon}
\newcommand{\sig}[0]{\sigma}
\newcommand{\se}[1]{\llbracket #1 \rrbracket}
\newcommand{\reff}[1]{Fig. \ref{#1}}
\newcommand{\refe}[1]{(\ref{#1})}

\usepackage{nomencl}
\makenomenclature

\nomenclature[zz_s]{$\bs\sig$}{Cauchy stress tensor}
\nomenclature[zz_s]{$\hat{\bs\sig}$}{effective Cauchy stress tensor}
\nomenclature[zz_s]{$\sig^\tx{eq}$}{equivalent von \txsc{Mises} stress}
\nomenclature[zz_s]{$\sig^\tx{y}$}{yield stress of plasticity model}
\nomenclature[zz_s]{$\sig^\tx{y,f}$}{yield stress at damage initiation}
\nomenclature[zz_e]{$\bs\eps$}{strain tensor}
\nomenclature[zz_e]{$\bs\eps^\tx{el}$}{elastic part of strain tensor}
\nomenclature[zz_e]{$\bs\eps^\tx{pl}$}{plastic part of strain tensor}
\nomenclature[zz_e]{$\overline\eps^\tx{pl}$}{equivalent plastic strain}
\nomenclature[zz_e]{$\overline\eps^\tx{pl,0}$}{equivalent plastic strain at damage initiation}
\nomenclature[zz_e]{$\overline\eps^\tx{pl,f}$}{equivalent plastic strain at failure}
\nomenclature[zz_g]{$\Gamma$}{cohesive interface}
\nomenclature[zz_g]{$\partial\Omega_{\overline{\tx t}}$}{Neumann boundary}
\nomenclature[zz_g]{$\partial\Omega_{\overline{\tx u}}$}{Dirichlet boundary}
\nomenclature[zz_o]{$\Omega$}{domain}
\nomenclature[zz_o]{$\Omega_+$}{domain of material phase +}
\nomenclature[zz_o]{$\Omega_-$}{domain of material phase -}
\nomenclature[zz_n]{$\nu$}{\txsc{Poisson's} ratio}
\nomenclature[zz_k]{$\kappa$}{scalar multiplier in plastic evolution equation}
\nomenclature[zz_d]{$\delta_\tx m$}{mixed mode separation}
\nomenclature[zz_d]{$\delta^0_\tx m$}{mixed mode separation at damage initiation}
\nomenclature[zz_d]{$\delta^\tx f_\tx m$}{mixed mode separation at failure}
\nomenclature[zz_l]{$\lambda$}{horizontal expansion of idealized interface profile}
\nomenclature[zz_g]{$\gamma$}{horizontal expansion of modified triangular profile}

\nomenclature[n_m]{$(\cdot)^\tx m$}{values related to the microscale}
\nomenclature[n_m]{$(\cdot)^\tx M$}{values related to the macroscale}
\nomenclature[n_e]{$E$}{\txsc{Youngs's} modulus}
\nomenclature[n_b]{$b$}{width of microscale domain}
\nomenclature[n_h]{$h$}{height of microscale domain}
\nomenclature[n_n]{$\bt n$}{unit normal vector on domain boundary}
\nomenclature[n_n]{$\bt n_\tx c$}{unit normal vector on cohesive interface}
\nomenclature[n_s]{$\bt s_\tx c$}{unit tangential vector on cohesive interface}
\nomenclature[n_t]{$\bt t_\tx c$}{cohesive traction vector on interface}
\nomenclature[n_t]{$\bt t$}{traction vector on domain boundary}
\nomenclature[n_t]{$\overline{\bt t}$}{prescribed traction vector on domain boundary}
\nomenclature[n_u]{${\bt u}$}{displacement vector}
\nomenclature[n_u]{${\tilde{\bt u}^\tx m}$}{displacement fluctuation vector on microscale}
\nomenclature[n_u]{$\overline{\bt u}$}{prescribed displacement vector on domain boundary}
\nomenclature[n_u]{$\se{\bt u}$}{separation vector on interface}
\nomenclature[n_u]{$\overline u^\tx{pl}$}{equivalent plastic displacement}
\nomenclature[n_x]{${\bt x}$}{position vector}
\nomenclature[n_w]{${\delta w^\tx M_\tx c}$}{virtual work at point on $\Gamma^\tx M_\tx c$}
\nomenclature[n_w]{${\delta \overline{W^\tx m_\tx c}}$}{volume average of virtual work of microdomain}
\nomenclature[n_d]{$D_\tx{CDM}$}{scalar damage variable of continuum damage model}
\nomenclature[n_f]{$f$}{yield condition of plasticity model}
\nomenclature[n_g]{$\tx G_\tx c$}{critical energy release rate}
\nomenclature[n_d]{$D_\tx{CZM}$}{scalar damage variable of cohesive zone model}
\nomenclature[n_k]{$k_I$}{stiffness of cohesive zone model in $I=\tx n$ normal and $I=\tx s$ shear direction}
\nomenclature[n_t]{$t^0_I$}{strength of cohesive zone model in $I=\tx n$ normal and $I=\tx s$ shear direction}
\nomenclature[n_t]{$t^0_I$}{strength of cohesive zone model in $I=\tx n$ normal and $I=\tx s$ shear direction}
\nomenclature[n_g]{$G_I$}{energy release rate in mode $I=\tx{I,II}$}
\nomenclature[n_g]{$G_{\tx cI}$}{critical energy release rate in mode $I=\tx{I,II}$}
\nomenclature[n_a]{$A$}{vertical expansion of idealized interface profile}
\nomenclature[n_g]{$G^\tx{eff}_{\tx c}$}{effective critical energy}
\nomenclature[n_t]{$t^\tx{eff}_{\tx c}$}{effective strength}
\nomenclature[n_t]{$t^\tx{M}_{\tx c,i}$}{the i-th discrete effective traction value}
\nomenclature[n_u]{$\se u^\tx{M}_{\tx c,i}$}{the i-th discrete separation value}
\nomenclature[n_l]{$L$}{effective critical energy}
\nomenclature[n_r]{$R_\tx a$}{arithmetic average of height profile}
\nomenclature[n_c]{$C$}{autocorrelation function}
\nomenclature[n_h]{$h(x)$}{interface height profile}
\nomenclature[n_l]{$L$}{length of measured height profile}
\nomenclature[n_l]{$l_\tx{ACF}$}{correlation length}

\usepackage{etoolbox}
\makeatletter
\patchcmd{\ps@pprintTitle}
  {Preprint submitted to}
  {}
  {}{}
\makeatother

\newcommand\blfootnote[1]{%
  \begingroup
  \renewcommand\thefootnote{}\footnote{#1}%
  \addtocounter{footnote}{-1}%
  \endgroup
}

\begin{document}
\begin{frontmatter}

\title{Microscale simulation of adhesive and cohesive failure in rough interfaces}


\author[IFKM]{Franz Hirsch}
\author[IFKM,DCMS]{Markus K\"astner\corref{cor}}
\address[IFKM]{Institute of Solid Mechanics, TU Dresden, 01062 Dresden, Germany}
\address[DCMS]{Dresden Center for Computational Materials Science (DCMS), TU Dresden, 01062 Dresden, Germany}
\cortext[cor]{Corresponding author. Tel.: +49\,351\,46332656.}
\ead{markus.kaestner@tu-dresden.de}

\blfootnote{© 2017. This manuscript version is made available under the CC-BY-NC-ND 4.0 license}
\blfootnote{https://creativecommons.org/licenses/by-nc-nd/4.0/}

\begin{abstract}
Multi-material lightweight designs, e.g. the combination of aluminum with fiber-reinforced composites, are a key feature for the development of innovative and resource-efficient products. The connection properties of such bi-material interfaces are influenced by the geometric structure on different length scales. In this article a modeling strategy is presented to study the failure behavior of rough interfaces within a computational homogenization scheme. We study different local phenomena and their effects on the overall interface characteristics, e.g. the surface roughness and different local failure types as cohesive failure of the bulk material and adhesive failure of the local interface. Since there is a large separation in the length scales of the surface roughness, which is in the micrometer range, and conventional structural components, we employ a numerical homogenization approach to extract effective traction-separation laws to derive effective interface parameters. Adhesive interface failure is modeled by cohesive elements based on a traction-separation law and cohesive failure of the bulk material is described by an elastic-plastic model with progressive damage evolution.
\end{abstract}


\end{frontmatter}

\section{Introduction} 

\begin{figure}
  \centering
    \includegraphics{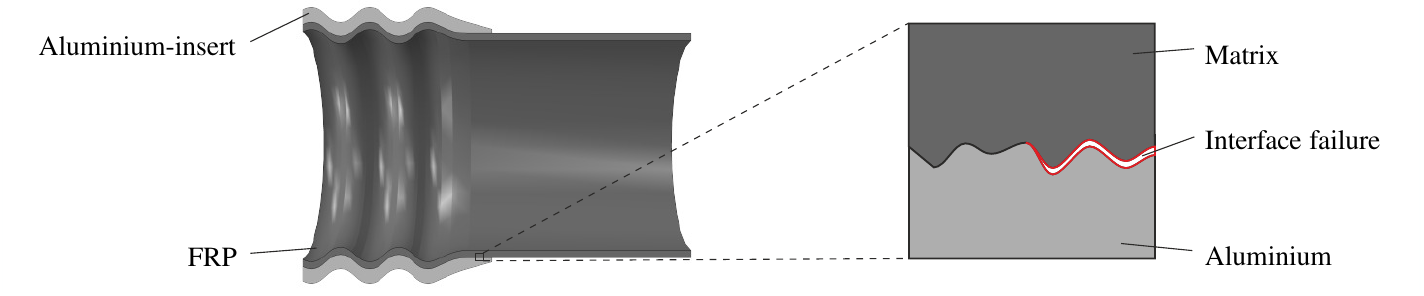}
\caption{Hybrid joint consisting of an aluminum insert and a fiber-reinforced plastic (FRP). Failure initiation of the bi-material interface between aluminum and matrix material is shown schematically.}
\label{fig:cad_intro}
\end{figure}

The application of fiber reinforced plastics (FRP) in multi-material lightweight structures requires innovative joining concepts to combine the advantages of FRP with conventional light\-weight materials as aluminum. Such hybrids are e.g. essential to realize appropriate load transfer elements in automotive or aircraft applications \cite{ashby2003,grujicic2008}. Different from technologies like bolted or riveted joints, which generally induce damage in the composite material, approaches that create the connection during the forming process itself instead of a subsequent joining step are promising solutions. Despite the intrinsic joining process, component failure often initializes in the bonding zone as shown in Fig. \ref{fig:cad_intro}, since the adhesive strength of polymer-metal interfaces is often much lower than the cohesive strength of the connected materials \cite{yao2002,zhao2007}. The adhesion between two materials can be influenced by different mechanisms. These include chemical/physical-chemical mechanisms (hydrogen and van der Waals forces), electrostatic mechanisms (difference in electrical charge) and mechanical mechanisms (interlocking) \cite{bruzzone2008}. This paper focuses on the latter mechanism by analyzing structured interfaces that create a mechanical interlock. For instance in \reff{fig:cad_intro}, the design of a macroscopic waviness or an increased microscopic interface roughness due to a certain pre-treatment of the aluminum surface may improve the mechanical properties of the joint.\\

\pagebreak
{\scriptsize\printnomenclature}
\pagebreak

A range of publications deal with the experimental investigation of structured interfaces and the resulting interface properties \cite{yao2002,lucchetta2011,kim2010,Cordisco2016}. In this context mainly microscopic structures have been studied. Lucchetta et al. \cite{lucchetta2011} investigated the correlation between the roughness of an aluminum-polymer interface and the resulting strength in lap shear tests. An aluminum plate was shot-peened with granular corundum to increase the surface roughness and over-molded by a polymer with/without fiber reinforcement. They concluded an enhanced interlocking, influenced by the interface roughness, fraction of reinforcing fibers and substrate temperature. In contrast to statistically random surface profiles, Kim et al. \cite{kim2010} investigated a polymer-metal interface with a designed micro-pattern. They used the end-notched flexure and single-leg bend test to provide pure mode II and mixed mode conditions. They show, that a strength increase can be obtained due to an increased roughness, if it causes a transition from local adhesive to cohesive failure.

Numerical models to describe and predict the failure behavior of bi-material interfaces have been proposed in e.g. \cite{yao2002,noijen2009,Zavattieri2007,Kaestner2016}. Yao and Qu \cite{yao2002} developed a fracture mechanics model to predict the amount of adhesive and cohesive failure in rough polymer-metal interfaces. These investigations were extended to a numerical fracture mechanics model by Noijen et al. \cite{noijen2009} and applied to an aluminum-epoxy interface. Pure adhesive mode I failure of sinusoidal structured interfaces were investigated numerically by Zavattieri et al. for identical elastic \cite{Zavattieri2007} and elastic-plastic \cite{Zavattieri2008} materials. Cordisco et al. extended this work to a sinusoidal interface between two elastically dissimilar \cite{Cordisco2012} and elastic-plastic dissimilar \cite{Cordisco2014} materials.

\begin{figure}
  \centering
  \begin{footnotesize}
  \includegraphics{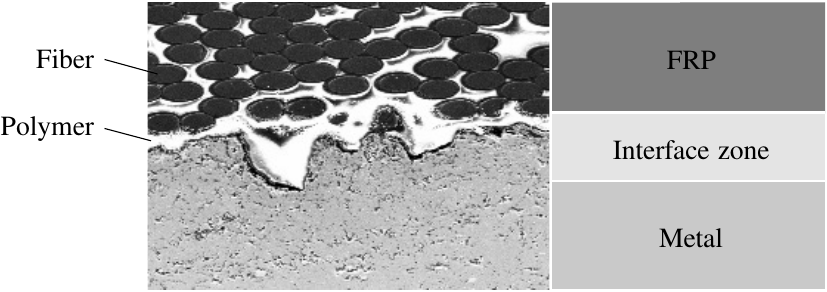}
  \end{footnotesize}
\caption{Microsection of the interface zone of a metal-composite hybrid.}
\label{fig:rem_scan}
\end{figure}

In the following, we consider the interface failure behavior on the microscale and the influence of local inhomogeneities. The microsection of the interface zone between a metal component and an FRP is shown in \reff{fig:rem_scan}, which indicates different inhomogeneities, e.g. the rough metal-polymer interface and embedded carbon fibers. The structure can be divided into a homogeneous  metal zone, an interface zone with pure polymer material which fills out the rough metal surface and an FRP zone. Subsequently the metal and interface zone is examined. Since there is a large separation of length scales, numerical multiscale simulation techniques are a suitable means to investigate the local phenomena in the vicinity of the interface and to predict effective interface properties. Recently, homogenization schemes for thin layers and interfaces, that provide a macroscopic traction-separation relationship instead of a stress-strain relation, have been developed. Matou\v{s} et al. \cite{Matous2008} proposed a multiscale cohesive approach to investigate thin inhomogeneous adhesive layers with a local continuum damage model. Similar homogenization schemes were formulated for adhesive layers by Alfaro et al. \cite{Alfaro2010} and for general adhesive/cohesive failure in quasi brittle solids by Verhoosel et al. \cite{Verhoosel2010}, where local damage was modeled with cohesive zone elements. 

In this contribution the multiscale formulation proposed by Alfaro et al. \cite{Alfaro2010} is adopted to study polymer-metal interfaces with a certain interface roughness. The description of the local material structure is based on the finite element method 
and the combination of a cohesive zone approach with continuum damage mechanics. This allows for the consideration of interface failure and damage evolution in the bulk material. The structure of the paper is as follows:
Section \ref{sec:numerical_models} provides a detailed description of the proposed modeling strategy including a review of the used homogenization scheme, a summary of the used constitutive models and a geometric description of the rough interface. 
This modeling approach is then used to study the general failure behavior and the interactions of different local phenomena for tension and shear loadings in Section \ref{Results}.

\section{Numerical modeling strategy}
\label{sec:numerical_models}

In this section the used approach to multiscale interface modeling is presented, i.e. bulk and layer homogenization techniques are applied to a representative interface section of a bi-material joint. After a summary of the constitutive models used within the homogenization approach, the geometrical description of interface profiles is addressed.

\subsection{Homogenization}

The objective of the presented homogenization approach is the prediction of an effective traction-separation relation for a  macroscopically flat interface which has to account for the effects of microscopic interface roughness and the interaction of local adhesive interfacial fracture and cohesive failure of the bulk material in the vicinity of the interface. The homogenization scheme used for the finite element simulations in Section \ref{Results} was proposed by Alfaro et al. \cite{Alfaro2010} and was originally applied to homogenize thin inhomogeneous adhesive layers with defined layer thickness. First applications to interfaces or layers of indefinite thickness can be found in e.g. \cite{Kaestner2016} and \cite{Palmieri2014}. 

\subsubsection{Boundary value problems}

\begin{figure}
  \centering
  \begin{footnotesize}
  \includegraphics{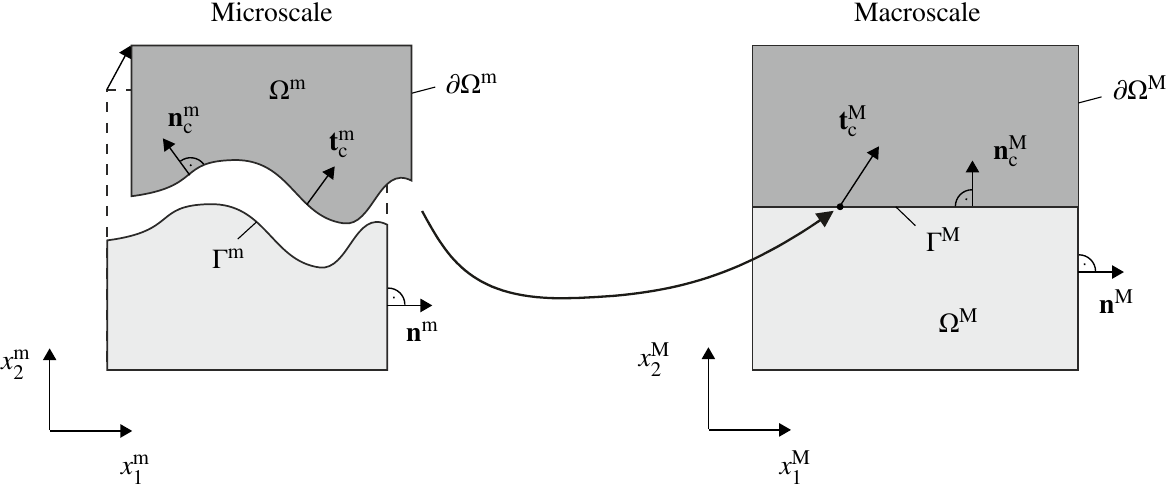}
  \end{footnotesize}
\caption{Idealized macroscopic and microscopic domains with a structured material interface.}
\label{fig:homog_scheme}
\end{figure}

In accordance with conventional multiscale analysis, we distinguish between the macroscale and the meso- or microscale with their individual boundary value problems for the two-dimensional case, \reff{fig:homog_scheme}. All quantities related to the macroscale are indicated with $(.)^\tx M$ and all microscopic quantities are labeled with $(.)^\tx m$. Without body forces, the quasi-static equilibrium is governed by the momentum balance and corresponding boundary and jump conditions on the macroscale

\begin{align}
 \tx{div}\left( \bs\sig^\tx M \right)&=0 & \tx{in }\Omega^\tx M\\
 \bs\sig^\tx M \cdot \bt n^\tx M_\tx c&= \bt t^\tx M_\tx c \left(\se{\bt u^\tx M}\right) & \tx{on }\Gamma^\tx M\\
 \bs\sig^\tx M \cdot \bt n^\tx M&= \overline{\bt t}^\tx M & \tx{on }\partial\Omega^\tx M_{\overline{\tx t}}\\
 \bt u^\tx M&=\overline{\bt u}^\tx M & \tx{on }\partial\Omega^\tx M_{\overline{\tx u}},
\end{align}

where $\bs\sig^\tx M$ represents the symmetric stress tensor in the domain $\Omega^\tx M$ and $\bt n^\tx M$ the unit normal vector on the boundary $\partial\Omega^\tx M=\partial\Omega^\tx M_{\overline{\tx t}}\cup\partial\Omega^\tx M_{\overline{\tx u}}$. Traction boundary conditions $\overline{\bt t}^\tx M$ and displacement boundary conditions $\overline{\bt u}^\tx M$ are prescribed to the parts $\partial\Omega^\tx M_{\overline{\tx t}}$ and $\partial\Omega^\tx M_{\overline{\tx u}}$ of the boundary, respectively . Cohesive tractions $\bt t^\tx M_\tx c$, that occur at the internal material interface on the macroscale, are governed by the effective separations $\se{\bt u}^\tx M$. These separations represent the jump of the displacement vector $\bt u^\tx M$ across the interface $\Gamma^\tx M$ and $\bt n^\tx M_\tx c$ is the corresponding unit normal vector.

The governing equations of the microscopic field problem read

\begin{align}
 \tx{div}\left( \bs\sig^\tx m \right)&=0 & \tx{in }\Omega^\tx m\\
 \bs\sig^\tx m \cdot \bt n^\tx m_\tx c&= \bt t^\tx m_\tx c & \tx{on }\Gamma^\tx m\\
 \bt u^\tx m&=\overline{\bt u}^\tx m & \tx{on }\partial\Omega^\tx m_{\overline{\tx u}}
\end{align}

where the traction boundary conditions were omitted in accordance with the used displacement boundary conditions defined later. An equivalent weak form of the boundary value problem, which is the basis for a finite element discretization as well as the formulation of equivalence criteria, is given by the principle of virtual work

\begin{equation}
\label{eq:virtual_work}
 \int_{\Omega^\tx m} \bs\sig^\tx m : \delta\bs\eps^\tx m \tx dV+\int_{\Gamma^\tx m}\bt t^\tx m_\tx c\cdot \delta\se{\bt u^\tx m} \tx dS = \int_{\partial\Omega^\tx m}\bt t^\tx m \delta\bt u^\tx m \tx dS,
\end{equation}

with the jump operator $\se{\bt a}=\bt a^+-\bt a^-$ and the infinitesimal strain tensor $\bs\eps^\tx m$.

\subsubsection{Boundary conditions of the micro domain}

\begin{figure}
  \centering
  \begin{footnotesize}
  \includegraphics{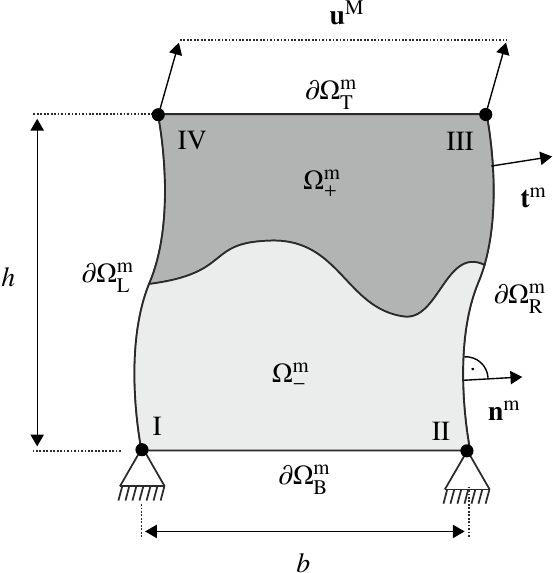}
  \end{footnotesize}
\caption{Microscopic domain with associated boundary conditions.}
\label{fig:rve}
\end{figure}

For the investigation of phenomena on the microscale, a suitable volume element (RVE) and admissible boundary conditions have to be identified to represent the effective interface behavior. In this  contribution the rectangular domain of \reff{fig:rve} will be used. Two material phases $\Omega^\tx m_+ \cup \Omega^\tx m_-=\Omega^\tx m$ are separated by a horizontal material interface $\Gamma^\tx m_\tx c$. The shape of the interface exhibits a roughness profile, which may be random or fully periodic with recurring structural elements. The microscopic displacement field

\begin{equation}
\label{eq:displ_micro}
 \bt u^\tx m=\bt u^\tx M +\tilde{\bt u}^\tx m
\end{equation}

can be decomposed into the macroscopic displacement $\bt u^\tx M$ and a small fluctuation $\tilde{\bt u}^\tx m$, induced by the local inhomogeneities. In sense of a representative segment of the interface it is assumed, that the micro domain in \reff{fig:rve} represents a volume element containing all main characteristics. Therefore, it could be repeated periodically in the $x_1$-direction. In  \reff{fig:rve} four corner nodes were introduced to relate the displacements on the microscale $\bt u^\tx m_I$, $I=\tx{I, II, III, IV}$ to the macroscopic separation $\se{\bt u^\tx M}$ of the interface $\Gamma^\tx M$:

\begin{equation}
\label{eq:hybrid_bc_1}
 \se{\bt u^\tx M}= \bt u^\tx m_\tx{IV}-\bt u^\tx m_\tx{I}=\bt u^\tx m_\tx{III}-\bt u^\tx m_\tx{II}.
\end{equation}

At these corner nodes the micro fluctuations $\tilde{\bt u}^\tx m_I=0$, $I=\tx{I, II, III, IV}$ must vanish in Equation \refe{eq:displ_micro}. To account for the periodicity of the interface in the $x_1$-direction, hybrid boundary conditions, i.e. a combination of periodic and linear displacement boundary conditions \cite{hirschberger2009}, are used. Consequently, periodic displacement and anti-periodic traction boundary conditions 

\begin{equation}
\label{eq:hybrid_bc_2}
 \tilde{\bt u}^\tx m_\tx L (s)= \tilde{\bt u}^\tx m_\tx R (s), \qquad \bt t^\tx m_\tx L(s)=-\bt t^\tx m_\tx R(s),
\end{equation}

with $s$ the local coordinate along the left and right RVE edges $\partial\Omega^\tx m_\tx L$ and $\partial\Omega^\tx m_\tx R$ are applied. Periodicity cannot be assumed in $x_2$-direction due to the transition from one macroscopic material phase into another. 
Therefore, vanishing fluctuation fields of the top and bottom edge $\partial\Omega^\tx m_\tx T$ and $\partial\Omega^\tx m_\tx B$

\begin{equation}
\label{eq:hybrid_bc_3}
 \tilde{\bt u}^\tx m_\tx T=\tilde{\bt u}^\tx m_\tx B=\bs 0
\end{equation}

result in linear displacement boundary conditions for these edges. Finally, the macroscopic separation can be related to the displacement of one corner node for a fixed bottom edge as shown in \reff{fig:rve}, i.e.

\begin{equation}
\label{eq:hybrid_bc_4}
 \se{\bt u^\tx M}= \bt u^\tx m_\tx{IV}=\bt u^\tx m_\tx{III}.
\end{equation}

In a displacement controlled homogenization scheme, the macroscopic deformation (separation) can be applied to the RVE immediately as a displacement prescribed to the top edge.

\subsection{Effective macroscopic traction}
\label{sec:effective_traction}

In addition to the connection between local displacement fields and the effective macroscopic interface deformation, similar expressions are necessary to link the tractions on both scales. Based on \txsc{Hills}'s averaging principle \cite{hill1967}, an appropriate criterion for interfaces can be formulated \cite{Alfaro2010}, as the equivalence of the virtual work density $\delta w^\tx M_\tx c$ at a single point on the macroscopic interface $\Gamma^\tx M$ and the volume average of the total virtual work $\overline{\delta W^\tx m}$ of the micro domain, i.e.

\begin{equation}
\label{eq:hill}
 \delta w^\tx M_\tx c = \overline{\delta W^\tx m}.
\end{equation}

The averaging procedure is denoted with an over-bar. In case of interfaces, the work conjugated quantities forming the virtual work of the macroscale

\begin{equation}
 \delta w^\tx M_\tx c=\bt t^\tx M_\tx c \cdot \delta\se{\bt u^\tx M}
\end{equation}

are the cohesive tractions $\bt t^\tx M_\tx c$ and the corresponding separations $\se{\bt u^\tx M}$. The microscopic volume average results from the right-hand side of Equation \refe{eq:virtual_work}:

\begin{equation}
\overline{\delta W^\tx m}=\frac{1}{b}\int\limits_{\Gamma^\tx m} \bt t^\tx m \cdot \delta \bt u^\tx m \tx dS,
\end{equation}

with the width $b$ of the micro domain. Now the displacement definition \refe{eq:displ_micro} in combination with the hybrid boundary conditions \refe{eq:hybrid_bc_2}-\refe{eq:hybrid_bc_4} can be used to evaluate the energy criterion \refe{eq:hill} in order to determine the macroscopic tractions $\bt t^\tx M_\tx c$. A detailed description of every transformation step can be found in \cite{Alfaro2010} or in the appendix of \cite{Palmieri2014}. The final expression of the macroscopic traction in terms of the microscale quantities reads 

\begin{equation}
\label{eq:effective_traction}
\bt{t}^\tx M_\tx c=\frac{1}{b}\int\limits_{\Gamma^\tx m_\tx T} \bt t^\tx m_\tx T \tx dS.
\end{equation}

With Equations \refe{eq:hybrid_bc_4} and \refe{eq:effective_traction} it is possible to extract the desired effective traction-separation law (TSL) $\bt{t}^\tx M_\tx c \left( \se{\bt u^\tx M} \right)$ from the analysis of the mechanical behavior in the microscale RVE. The proposed scheme was used by Alfaro et al. \cite{Alfaro2010} to homogenize adhesive layers of constant thickness. In that case, the domain height $h$ could be chosen equal to the layer thickness and the width $b$ was considered in the homogenization step, cf. Equation \refe{eq:effective_traction}. For interface investigations there is no determined height characterizing the macroscopic interface. A suitable choice for the height of the RVE has to ensure that the failure localization can fully develop within the RVE \cite{vossen2014}.

\subsection{Constitutive models}

In this section all material and interface models are summarized. Since all considered quantities belong to the microscale, the corresponding marker $(.)^\tx m$  is omitted for brevity of the notation. The models describe a general metal-polymer interface as considered in Section \ref{sec:simulation_tension}. Regarding the local failure mechanisms, we distinguish between cohesive and adhesive failure. On the macroscale the  interface behavior is described by an effective TSL. Therefor the failure of these interfaces is called adhesive failure. Nevertheless, failure on the microscale can be a combination of cohesive failure of the bulk material and adhesive failure of the local interface. The former is accounted in terms of a continuum damage model (CDM) in Section \ref{ssec:CDM} and the latter is modeled by a cohesive zone model (CZM), Section \ref{ssec:CZM}.

\subsubsection{Linear elastic model}

Since the strength of the metal component is assumed to be much higher than the polymer strength, the behavior is modeled as isothermal linear elastic material 

\begin{equation}
 \bs\sig={}^4\bt C : \bs\eps, \qquad C_{ijkl}=2\mu\delta_{ik}\delta_{jl}+\lambda\delta_{ij}\delta_{kl}
\end{equation}

with the \txsc{Lam\'e} parameters $\mu =E/(2(1+\nu))$ and $\lambda = E/((1+\nu)(1-2\nu))$ depending on the \txsc{Young}'s modulus $E$ and the \txsc{Poisson}'s ratio $\nu$.

\subsubsection{Ductile damage model}
\label{ssec:CDM}

\begin{figure}
  \centering
  \begin{footnotesize}
  \centering
   \includegraphics{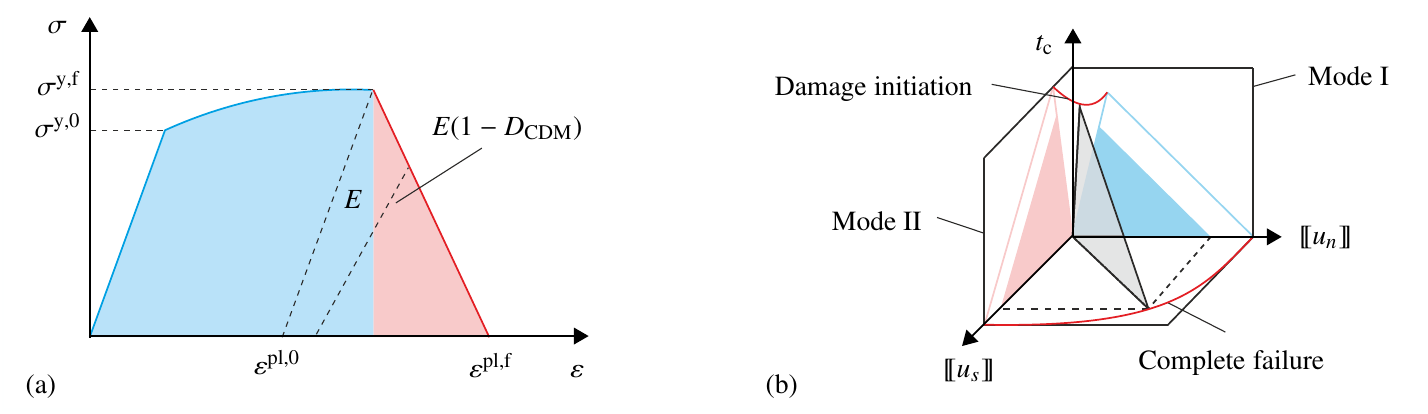}
  \end{footnotesize}
\caption{Schematic representation of a) the elastic-plastic damage model and b) the bilinear traction-separation law used for cohesive zone elements.}
\label{fig:cdm_tsl}
\end{figure}

To describe the cohesive failure within the polymer material, a von \txsc{Mises} plasticity model is used in combination with a progressive damage evolution. This model generically represents the inelastic behavior of a typical polymer and enables the study of the interaction between the bulk deformation and the roughness of the corresponding interface. The model definition starts with the basic assumption of the decomposition of the total strain $\bs\eps$ into an elastic part $\bs \eps^\tx{el}$ and plastic part $\bs\eps^\tx{pl}$: 

\begin{equation}
 \bs\eps=\bs\eps^\tx{el}+\bs\eps^\tx{pl}
\end{equation}
The stress response of the material is given by
\begin{equation}
 \hat{\bs \sigma} = {}^4\bt C : (\bs\varepsilon-\bs\varepsilon^\tx{pl}),
\end{equation}

with the effective stress tensor $\hat{\bs \sig}$ acting in the undamaged material. The effective stress tensor is introduced within the concept of effective stress \cite{Kachanov1985} to consider isotropic damage in sense of classical continuum damage mechanics. Therefor the nominal stress $\bs\sig$ of a damaged continuum is related to the effective stress $\hat{\bs\sig}$ in terms of

\begin{equation}
 \bs\sig=(1-D_\tx{CDM})\hat{\bs\sig}
\end{equation}

with $D_\tx{CDM}$ a nondecreasing scalar damage variable.  The damage variable $D_\tx{CDM}$ ranges between $0$ corresponding to the undamaged material and $1$, where the material is fully damaged and no stresses are transmitted any more. The evolution of the plastic strain $\bs\eps^\tx{pl}$ is governed by the definition of the flow rule and the yield condition. The yield condition $f$ separates the elastic domain $f<0$, where the response is purely elastic with no plastic yielding from the critical states $f=0$, where a change in $\bs\eps^\tx{pl}$ is possible.  The yield condition in case of isotropic von \txsc{Mises} plasticity takes the form

\begin{equation}
\label{eq:yield_condition}
 f=\sig^\tx{eq}-\sig^\tx y \left( \overline\eps^\tx{pl} \right) \leq 0,
\end{equation}

with the scalar equivalent von (\txsc{Mises}) stress $\sig^\tx{eq}$ and the yield stress $\sig^\tx y$. The equivalent stress $ \sig^\tx{eq}=\sqrt{\frac{3}{2}\bt s : \bt s}$ depends only on the deviatoric part of the stress tensor $\bt s=\hat{\bs\sig}-\frac{1}{3}tr(\hat{\bs\sig})$. A similar scalar measure for the plastic deformation history is the equivalent plastic strain $\overline\eps^\tx{pl}=\int_0^t\sqrt{\frac{2}{3} \dot{\bs\eps}^\tx{pl} : \dot{\bs\eps}^\tx{pl}} \tx d\tau$, which is used as driving variable within the hardening law $\sig^\tx y \left( \overline\eps^\tx{pl} \right)$. The conventional flow rule of von \txsc{Mises} plasticity with isotropic hardening reads

\begin{equation}
 \dot{\bs\eps}^\tx{pl}=\kappa\frac{3}{2}\frac{\bt s}{\sig^\tx{eq}},
\end{equation}

based on the normality rule and a scalar multiplier $\kappa$. An additional evolution equation is necessary to describe the progressive damage process driven by plastic deformation. A simple linear degradation path was chosen, determined by the evolution equation

\begin{equation}
 \dot D_\tx{CDM}=\frac{\sig^\tx{y,f}}{2 G_\tx f} \dot{\overline u}^\tx{pl},
\end{equation}

with $\sig^\tx{y,f}$ the equivalent stress at damage initiation and $G_\tx f$ the energy release rate. It is well known, that stress-strain relationships in CDM introduce an undesired mesh dependence. If the mesh is refined, strain localization effects will cause an unphysical decrease in the dissipated energy. Therefor the  equivalent plastic displacement $\overline u^\tx{pl}$ was introduced rather than the plastic strain itself to reduce the mesh dependence. This modification based on \txsc{Hillerborg} et al. \cite{Hillerborg1976} uses a characteristic element length to replace the equivalent plastic strain with the corresponding displacement.\\
The uniaxial stress-strain relationship of the presented model is shown in \reff{fig:cdm_tsl}~a). After a linear slope, the material starts to yield at $\sig^\tx{y,0}$, followed by the hardening regime $\sig^\tx y \left( \overline\eps^\tx{pl} \right)$. At $\eps^\tx{pl,0}$ damage initiates and a linear softening regime follows, where the initial elastic constant $E$ is degraded up to full material failure at $\eps^\tx{pl,f}$.

\subsubsection{Interface model}
\label{ssec:CZM}

The inelastic behavior of the interface is described in terms of a cohesive zone model (CZM), i.e. the deformation behavior in the process zone ahead of the crack tip is determined by a TSL. The elementary characteristics of most TSL are determined by the strength, the critical energy release rate and a certain model shape. Often no significant effects of the TSL shape on the overall mechanical behavior are expected \cite{Valoroso2006}. Hence, the bilinear TSL proposed by Camanho et al. \cite{Camanho2003}

\begin{equation}
\bt t_\tx c=(1-D_\tx{CZM})\left( k_\tx n \langle \se{u_\tx n}\rangle_+ \bt n_\tx c + k_\tx s  \se{u_\tx s} \bt s_\tx c\right)+k_\tx n\langle \se{u_\tx n}\rangle_- \bt n_\tx c
\end{equation}

is used, with the elastic stiffness $k_I$ and separations $\se{u_I}$ in normal $I=\tx n$ and shear direction $I=\tx s$, respectively. The normal and shear unit vectors $\bt n_\tx c$ and $\bt s_\tx c$ correspond to the local interface orientation. The  \txsc{Macaulay} brackets $\langle x \rangle_\pm=1/2 (x \pm |x|)$ ensure, that the damage variable $D_\tx{CZM}$ only affects the normal stiffness under tension loading to prevent interface penetration under compression loading. Similar to the damage variable in the CDM, the scalar damage variable $D_\tx{CZM}$ ranges between $0$ indicating an intact interface and $1$, where the interface is fully debonded. The schematic representation of the TSL is shown in Fig. \ref{fig:cdm_tsl}~b). Damage is initialized after a linear slope and triggered by the quadratic stress criterion

\begin{equation}
 \left( \frac{\langle t_\tx n \rangle}{t^0_\tx n} \right)^2 + \left( \frac{ t_\tx s}{t^0_\tx s} \right)^2 -1=0,
\end{equation}

with the traction coordinates $t_I$ and the cohesive strengths $t_I^0$. As indicated by the \txsc{Macaulay} brackets, damage initiates only under tension and shear conditions. Afterwards, a linear decrease of the elastic stiffness $k_I$ follows up to full failure determined by a power law criterion

\begin{equation}
 \left( \frac{G_\tx I}{G_\tx{cI}} \right)^\alpha + \left( \frac{G_\tx{II}}{G_\tx{cII}} \right)^\alpha -1=0,
\end{equation}

with the pure mode energy release rates $G_I$, the critical energy release rates $G_{\tx cI},~I=\tx{I,II}$ and the exponent $\alpha$ to weight the mixed mode behavior. All model parameters in normal and shear direction correspond to pure mode I and mode II failure of the interface. To describe the damage evolution under general loading conditions, the mixed mode separation $\delta_\tx m=\sqrt{\langle \se{u_\tx n} \rangle^2+\se{u_s}^2}$ acts as damage driving value within the evolution equation

\begin{equation}
 D_\tx{CZM}=\frac{\delta^\tx f_\tx m (\delta^\tx{max}_\tx m -\delta^0_\tx m)}{\delta^\tx{max}_\tx m(\delta^\tx f_\tx m-\delta^0_\tx m)},\qquad \delta^\tx{max}_\tx m = \underset{\tau \leq t}{\tx{max}}[\delta_\tx m(\tau)],
\end{equation}

with the mixed mode separation at damage onset $\delta^0_\tx m$ and the mixed mode separation $\delta^\tx f_\tx m$ at the fully debonded state.

\subsection{Characterization of rough interfaces}

In addition to constitutive properties, the investigation of rough interfaces or surfaces requires suitable measures to quantify and compare different geometric characteristics. The notion roughness implies a deviation from an ideal or nominal surface with characteristic dimensions in the micro or nano length scales. These deviations may be periodic or random. Periodic, structured interfaces are created on the microscale and mesoscale during cutting processes for instance and are determined by the geometry and process parameters of certain cutting tools. However, natural surfaces or pre-treatments like sandblasting usually result in random height profiles. There exist numerous definitions of characteristic values to describe different properties of the surface, where the significance often varies from application to application. In this contribution the notion roughness or roughness ratio concerns the ratio of two characteristic values describing the horizontal and vertical expansion of the deviations. 

\subsubsection{Random surfaces}

\begin{figure}
  \centering
  \begin{footnotesize}
   \includegraphics{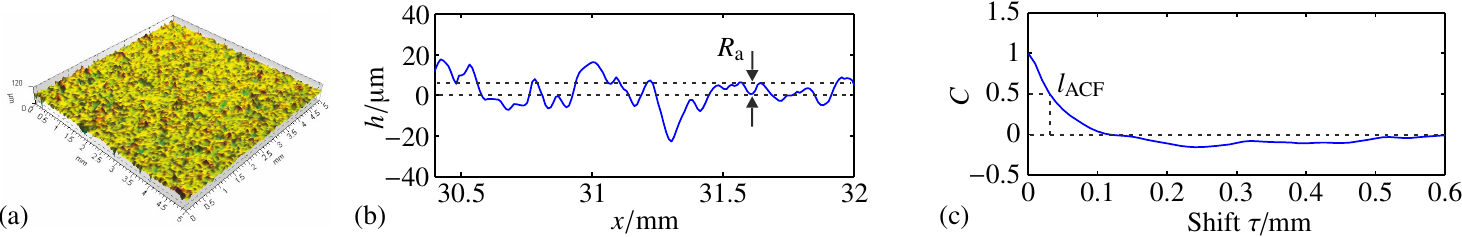}
  \end{footnotesize}
\caption{Characteristic length of random surfaces: a) Surface measurement with 3D laser profilometer, b) extracted 2D height profile $h(x)$, c) autocorrelation function of 2D height profile.}
\label{fig:surf_charact}
\end{figure}

An example for a statistically random surface is shown in \reff{fig:surf_charact}~a), where a laser profilometer was used to extract a 3D height profile from a sandblasted aluminum surface. The extraction of the 2D height profile as one line across the surface is shown in \reff{fig:surf_charact}~b). This profile can be used to extract the relevant  geometric characteristics. Yao et al. \cite{yao2002} used for instance  the arithmetic average 

\begin{equation}
 R_\tx a=\frac{1}{L}\int_0^L |h(x)| \tx dx
\end{equation}

to characterize the height and the autocorrelation length $l_\tx{ACF}$ to characterize the width of a certain height profile over a sampling length $L$. Hence, they formulated the non-dimensional roughness ratio $R_\tx a/l_\tx{ACF}$. To determine $l_\tx{ACF}$, the autocorrelation function (ACF) of the height profile $h(x)$ was used and normalized by the standard deviation $\sigma$ :

\begin{equation}
 C(\tau)=\underset{L \rightarrow \infty}{\tx{lim}} \frac{1}{\sigma^2L}\int_0^L h(x)h(x+\tau) \tx dx
\end{equation}

The result of the ACF is shown in \reff{fig:surf_charact}~c). At the origin the ACF reaches $C=1$, indicating that the profile is completely correlated to itself. The increase of the profile shift $\tau$ to itself results in the decrease of the ACF value until the profile is completely uncorrelated $C=0$ . When the ACF falls below a critical value, the corresponding shift is defined as correlation length $l_\tx{ACF}$, which is a suitable measure of the average horizontal dimension of the structure. The definitions of the critical ACF value vary in literature, e.g. $C=0.5$ was used by Yao et al. \cite{yao2002}.  It is possible to generate such random rough surfaces for implementation into an FE model. Temizer et al. \cite{Temizer2011} presented a generalized random-field model to generate random rough surfaces with import option into an FE framework. To characterize the generated surfaces, they used the root-mean-square roughness and ACF for the average height and width, respectively. The local failure behavior of random surfaces is quite complex, nevertheless only an average response is often of interest and obtainable in experiments. Therefore, it is reasonable to analyze idealized profiles, where the key geometric characteristics were fitted to the average values of the random surface. This approach is followed below.

\subsubsection{Periodic structured surfaces}

\begin{figure}
  \centering
  \begin{footnotesize}
   \includegraphics{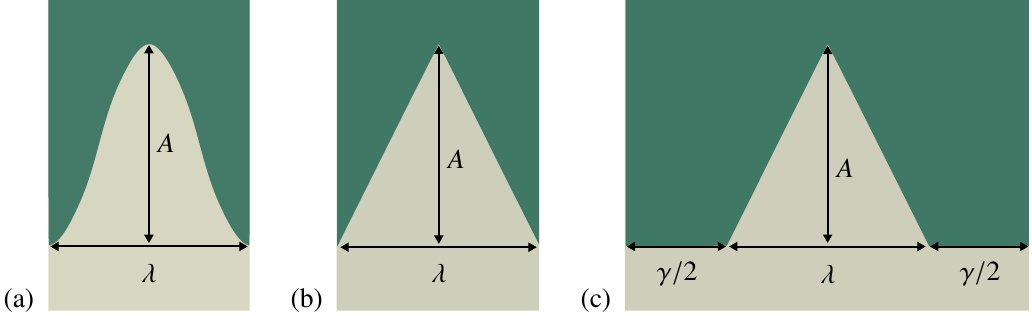}
  \end{footnotesize}
\caption{Idealized surface profiles: a) sinusoidal, b) triangular, c) modified triangular with flat area.}
\label{fig:ideal_profiles}
\end{figure}

Periodic, structured surfaces can either be geometries of a designed surface structure on the mesoscale or be representative of the characteristics of random surfaces. Some examples of idealized surfaces are shown in \reff{fig:ideal_profiles}, e.g. sinusoidal, triangular and triangular profiles with a certain flat part. Sinusoidal idealizations, \reff{fig:ideal_profiles}~a), were already chosen by Zavattieri et al. \cite{Zavattieri2007}.  The triangular interfaces, \reff{fig:ideal_profiles}~b) and c), are inspired by the investigations of Yao et al. \cite{yao2002}. Since the study in this work is attributed to the microscale, the triangular idealizations (jigsaw profile) appear more suitable because realistic roughness profiles of natural surfaces are often ragged with sharp corners and edges.

\section{Results and Discussion}
\label{Results}

 In this section the presented constitutive and geometric models are used in combination with the homogenization scheme to study the failure behavior of a bi-material interface with a certain surface roughness. The objective of this study is to analyze the influence of an increasing interface roughness on the resulting effective interface properties. Numerical simulations provide insight into the principal failure mechanisms and improve the understanding of the debonding process. A default configuration, i.e. a triangular profile with an initial roughness ratio of $A/\lambda=1$, \reff{fig:rve_tension_B}~a), is said to mimic a general aluminum-polymer interface with the following characteristics: The aluminum phase exhibits the highest strength of all interface components with comparatively high stiffness. Thus, no plastic yielding and damage are expected and the material is modeled linear elastic with $E_\tx{Al}=75~\tx{GPa}$ and $\nu_\tx{Al}=0.25$. In contrast, the polymer is very soft resulting in a mismatch of the elastic constants of $E_\tx P/E_\tx{ Al}=0.1$ and equal \txsc{Poission} ratios $\nu_\tx P=\nu_\tx{Al}=\nu$. Since we are interested in the adhesion enhancement of interfaces with a low surface binding by increasing its roughness, the yield point of the polymer $\sig^\tx{y,0}=3 t^0$ is chosen well above the interface strength. A very low hardening modulus of $0.2~\tx{GPa}$ defines the plastic behavior up to damage initiation at $\overline\eps^\tx{pl,0}=0.04$. The damage evolution is modeled in terms of a linear softening regime governed by the energy release rate of $G_\tx f=0.2~\tx{J}/\tx{mm}^2$. This generic material model represents a soft polymer with a distinct plastic regime and a rapidly evolving damage after the yield point which is a typical behavior of thermoplastic matrix systems. The adhesive interface properties were chosen equal in normal and shear direction, i.e. $t^0_\tx n=t^0_\tx s=t^0=25~\tx{MPa}$ with low critical fracture energies $G_\tx{cI}=G_\tx{cII}=G_\tx c=10~\tx J/\tx{mm}^2 $ and $\alpha=1$. Although, realistic interfaces may have different properties in normal and shear direction even for flat profiles on the microscopic scale, isotropic CZM parameters are chosen to investigate the effective strength enhancement as a result of pure geometric interface variation. As already mentioned in Section \ref{sec:effective_traction}, the RVE height is not determined, while the RVE width is chosen equal the width $\lambda$ of the idealized triangular profile. In this work, all simulations were performed with a constant domain height corresponding to the maximum height of the interface profile multiplied by a factor of $4.5$ to ensure that the failure localization can fully develop within the RVE \cite{vossen2014}. The notation of default configuration is referred to the generic set of parameters introduced above.

\subsection{Characteristic interface properties}

\begin{figure}
  \centering
  \begin{footnotesize}
  \includegraphics{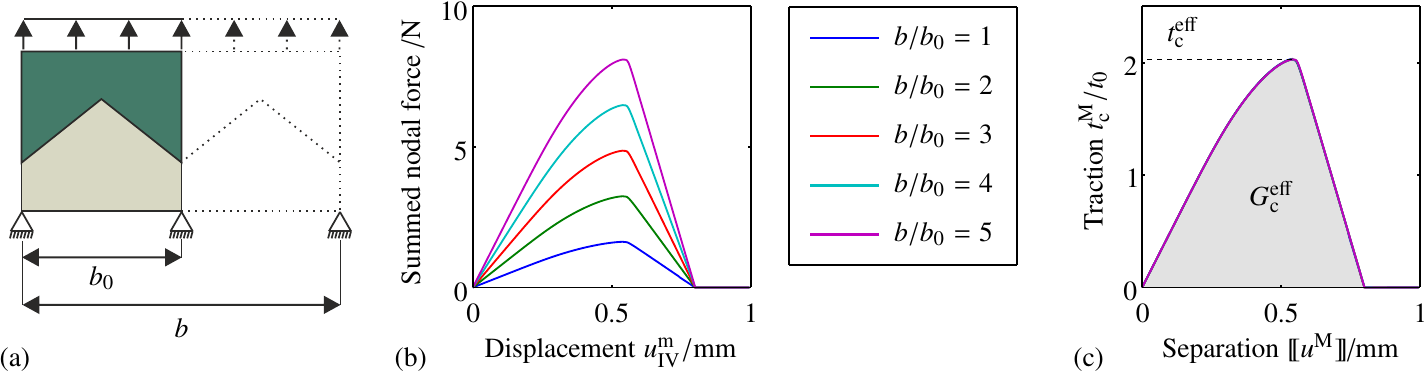}
  \end{footnotesize}
\caption{Simulation of mode I failure of a sinusoidal interface for different domain widths: a) Schematic representation of the RVE, b) resulting nodal force of the top edge and c) effective traction-separation law.}
\label{fig:homog_varWidth}
\end{figure}

The introduction of effective tractions in Equation \refe{eq:effective_traction} allows for the extraction of  width-independent effective TSLs. This independence is demonstrated by several simulation results for RVEs with different microscale domain sizes. In addition, the key parameters characterizing the effective adhesion properties are identified in this section. A micro model as shown in \reff{fig:homog_varWidth}~a) is used with an idealized interface profile. The RVE is loaded under tension conditions to simulate a mode I failure of the interface. Only adhesive failure is considered with the isotropic interface failure behavior of the default configuration. The materials near the interface show a mismatch of the elastic properties. The roughness ratio $A/\lambda$ was held constant, see \reff{fig:homog_varWidth}~a). The effective model response is shown in \reff{fig:homog_varWidth}~b), were the resulting nodal forces of the top edge are plotted versus the effective macroscopic separation defined by \refe{eq:hybrid_bc_4}. As expected, an increasing domain width results in higher maximum forces. After applying the proposed homogenization approach in \refe{eq:effective_traction}, coinciding effective TSL were obtained. After a linear slope, the response becomes nonlinear due to local debonding within the idealized interface. As the maximum is reached, the behavior of the complete interface is determined by the linear degradation of the cohesive zone model.\\ The response still depends on the domain height, since the homogenization is not performed in normal direction with respect to the interface. If the RVE size is not small compared to the macroscopic dimensions, the height could be considered as initial cohesive height on the macroscale. In this contribution, the domain height is held constant for all roughness ratios $A/\lambda$. The effective traction-separation curves can be used directly as macroscopic constitutive models within a coupled multiscale formulation, as input data for a subsequent curve fitting or to extract characteristic cohesive zone properties. In the following sections, the TSL is assembled by the extraction of discrete pairs of effective traction and separation values over the load history.  Characteristic properties are identified as the effective maximum traction

\begin{equation}
\label{eq:eff_strength}
 t^\tx{eff}_{\tx c}=\tx{max}(t^\tx M_{\tx c,i}), \qquad \tx{with } i=1,2,...,N
\end{equation}
and the effective critical energy 
\begin{equation}
\label{eq:eff_energy}
 G_\tx c^\tx{eff} = \sum_{i=1}^{N-1} \frac{1}{2}(t^\tx M_{\tx c,i}+t^\tx M_{\tx c,i+1}(\se{u^\tx M_{i+1}}-\se{u^\tx M_{i}}),
\end{equation}

with $N$ the number of discrete values of the effective TSL. 

\subsection{Simulation of interface failure under tensile loadings}
\label{sec:simulation_tension}

In general, the proposed numerical model can be loaded under arbitrary mixed mode conditions. The studied loading scenarios in this contribution include pure mode I and mode II conditions.  The former is presented in this section while the latter is discussed in Section \ref{sec:results_shear}. The investigated RVE in default configuration with corresponding loading conditions is shown in \reff{fig:rve_tension_B}~a). The roughness dimension on the microscale depends on the pretreatment of the metal surface. The microsection in \reff{fig:rem_scan} shows a sandblasted metal surface, that results in a high roughness ratio of the interface. Dependent on the blast pressure and the grain size this dimension can vary over a substantial range. Therefore, ratios $A/\lambda=0.25...1.4$ are analyzed in the following studies.  First, the general failure behavior is explained followed by the numerical analysis of microstructural model variations. 
 
\subsubsection{General failure behavior} 
 
\begin{figure}
  \centering
  \begin{footnotesize}
    \includegraphics{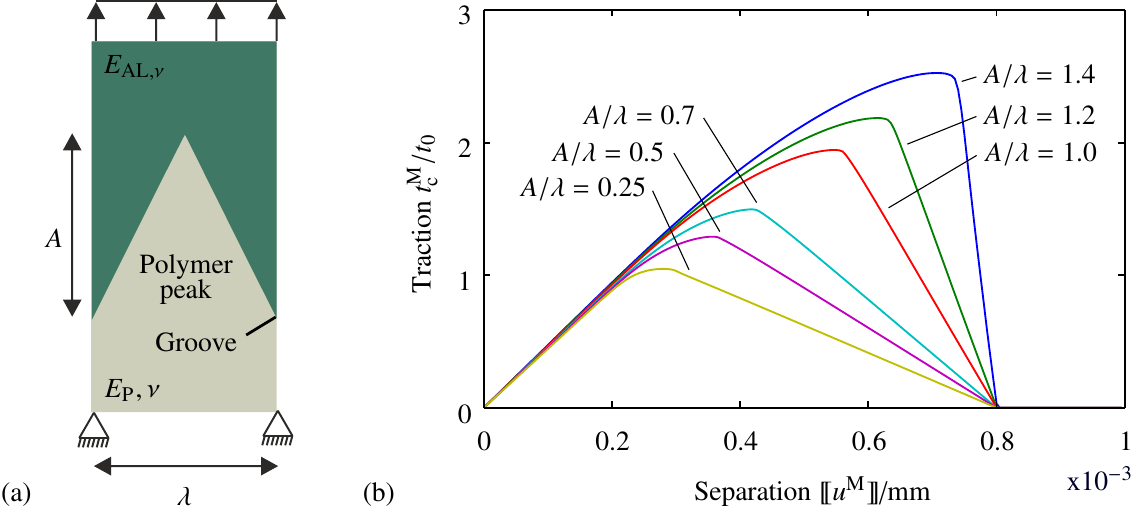}
  \end{footnotesize}
\caption{Macroscopic mode I loading of triangular interface: a) schematic RVE of the default configuration and tested loading conditions and b) the effective TSL for different roughness ratios $A/\lambda$.}
\label{fig:rve_tension_B}
\end{figure}

The effective traction-separation curves resulting from the homogenization scheme are shown in \reff{fig:rve_tension_B}~b). The results were normalized to the properties of the underlying CZM, representing the properties of a perfectly flat interface. With increasing roughness ratio $A/\lambda$, we observe an improvement of the adhesion properties of the surface due to the increase of the surface area. The change in the ratio $A/\lambda$ is performed for constant profile heights $A$ and varying $\lambda$. The nonlinear range between damage initiation and maximum traction results from the transition from initial local debonding of the interface up to the state, where all points of the interface reached the interface strength and are involved in the active damaging process. The degradation path of the post-critical range exhibit is governed by the linear evolution equation of the used CZM. As expected, the states of complete debonding coincide at a critical separation value determined by the microscopic TSL. 
 
 The evolution of debonding is shown in \reff{fig:strength_tension_B}~a). Due to the stress concentrations around the sharp corners of the triangular interface, interfacial damage initiates at the groove of the soft polymer material. With increasing load, the interface crack propagates along the straight surface edges to the profile peak. This pre-critical crack propagation causes the mentioned nonlinearity in the hardening regime of the effective TSL in \reff{fig:rve_tension_B}~b). The state of full debonding is characterized by a traction free material interface. 
 
\subsubsection{Influence of roughness enhancement} 
 
\begin{figure}
  \centering
  \begin{footnotesize}
  \centering
    \includegraphics{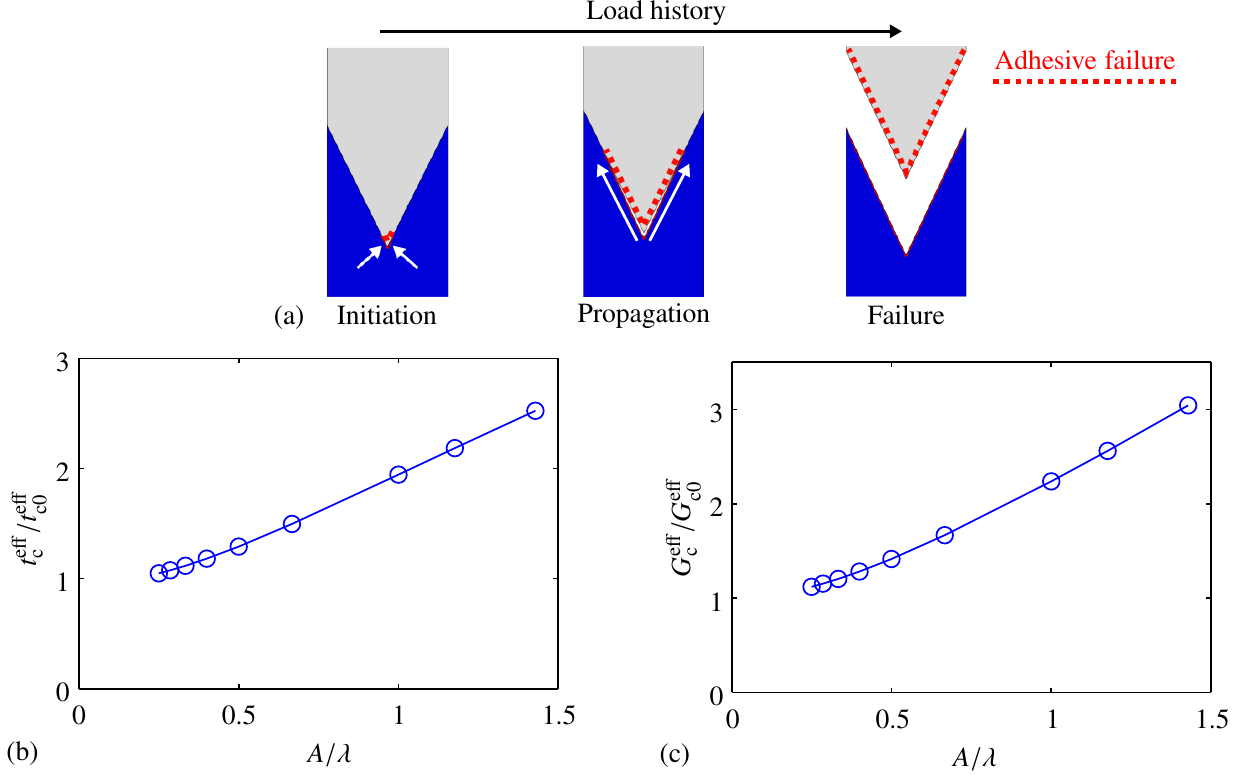}
  \end{footnotesize}

\caption{Macroscopic mode I loading of triangular interface in default configuration: a) Evolution of interface debonding during load history, b) effective strength and c) effective critical fracture energy.}
\label{fig:strength_tension_B}
\end{figure}

 The characteristic effective interface properties defined in Equation \refe{eq:eff_strength} and \refe{eq:eff_energy} are plotted versus the corresponding roughness ratio $A/\lambda$ in \reff{fig:strength_tension_B}~b) and \reff{fig:strength_tension_B}~c), respectively. Both curves were normalized with respect to the corresponding properties of the used CZM. At low ratios $A/\lambda$ the interface profile  and the effective properties converge to a flat interface with local CZM properties. Both quantities increase with rising roughness ratio, indicating an enhancement of the bi-material connection.

Since the interface strength is far below the polymer strength ($\sig^\tx{y,0}/t^0=3$), no bulk damage is initiated, even for high interface roughness ratios. Hence, the improvement of adhesive properties is caused by the increase of the contact area of the interface only. A shift from adhesive to cohesive failure can be observed only for lower ratios $\sig^\tx{y,0}/t^0=3$ or surface profiles with a mechanical interlock in vertical direction. 

\subsubsection{Influence of interface contour}
\label{sec:tension_influence_contour}

\begin{figure}
  \centering
  \begin{footnotesize}
    \includegraphics{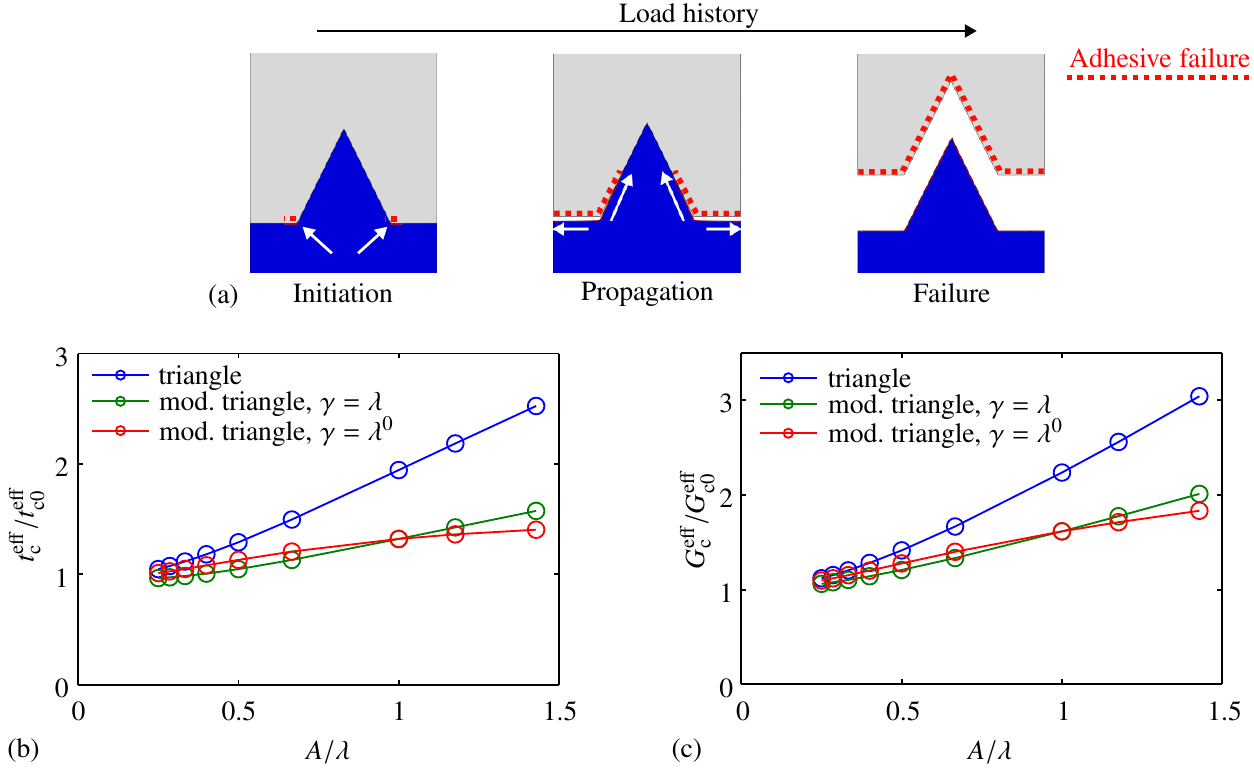}
  \end{footnotesize}
\caption{Macroscopic mode I loading different interface contours: a) Evolution of debonding of modified triangular interface with $A/\lambda=1$ during load history and b) effective strengths and  c) effective critical fracture energies.}
\label{fig:strength_tension_A}
\end{figure}

Below, the behaviour of the previously analyzed triangular interface geometry will be compared to a modified shape in order to investigate the influence of the profile. It is noted, that these idealized profiles with a defined roughness ratio $A/\lambda$ can be interpreted as an average of random rough surfaces on the microscale or a constructed surface on the mesoscale for instance. Different from sinusoidal profiles, these sharp-edged profiles cause damage inducing stress concentrations which is expected to be closer to naturally formed surfaces, e.g. metal surfaces after sandblasting.

A variation of the triangular interface profile is introduced in terms of flat domains in between every triangle, \reff{fig:ideal_profiles}~c). As a consequence, the profile symmetry was eliminated with respect to the horizontal direction. This profile is able to mimic certain skewness properties of the surface. In this investigation, two different choices for the additional interface parameter describing the width of the flat domain are considered. In the first case, it is equal to the actual width of the triangle $\gamma=\lambda$ while it is held constant $\gamma=\lambda^0$ in the second case. The influence of the choice of $\gamma$ is investigated in Section 3.3.3.

The evolution of interfacial debonding is shown in \reff{fig:strength_tension_A}~a) for a ratio $A/\lambda=1$. Similarly to the triangular profile, adhesive damage initiates at the groove of the soft polymer material due to stress concentrations. However, the crack propagates initially along the flat horizontal domain and interfacial failure of the edges of the profile peak is delayed. The qualitative shapes of the corresponding TSL are similar to the TSL of the triangular interface in \reff{fig:rve_tension_B}~b). 

The comparison of the effective strength and energy release rate is shown in \reff{fig:strength_tension_A}~b) and c), respectively. Over the complete roughness range, the triangular profile shows a better adhesion and higher sensitivity to an increase of the roughness ratio $ A/\lambda$. At low ratios $A/\lambda$, the strength and in particular the critical energies vary slightly between both interface versions. The effective properties of the modified interface with $\gamma=\lambda$ are below the properties of the interface with $\gamma=\lambda^0$ for $A/\lambda<1$, because flat part without increased surface area is constant and always smaller in a uniform interface section. For ratios $A/\lambda>1$ an inverse relation is observed following the same argumentation. In summary, this study of mode I failure already shows the significance of a continuously structured interface and the loss of adhesion due to flat domains with low roughness ratios $A/\lambda$.

\subsection{Simulation of interface failure under shear conditions}
\label{sec:results_shear}

For macroscopic mode I loading conditions, the increase of the roughness ratio $A/\lambda$ results in enhanced adhesion between the aluminum and the polymer material. Since the investigated default configuration showed no local cohesive failure over a wide range of $A/\lambda$, the principle mechanism that caused the mentioned improvement is a simple increase of the contact area. Under macroscopic mode II (shear) loading all interface profiles with $A/\lambda > 0$ form a mechanical interlock with different interactions between adhesive and cohesive failure on the microscale. The investigated RVE in default configuration with corresponding loading conditions is shown in \reff{fig:rve_shear_B}~a). Similar to the study of mode I failure, a general description of the damaging process of the default configuration is presented with subsequent variations of certain model parameters. 

\subsubsection{General failure behavior}
\label{ssec:shear}

\begin{figure}
  \centering
  \begin{footnotesize}
    \includegraphics{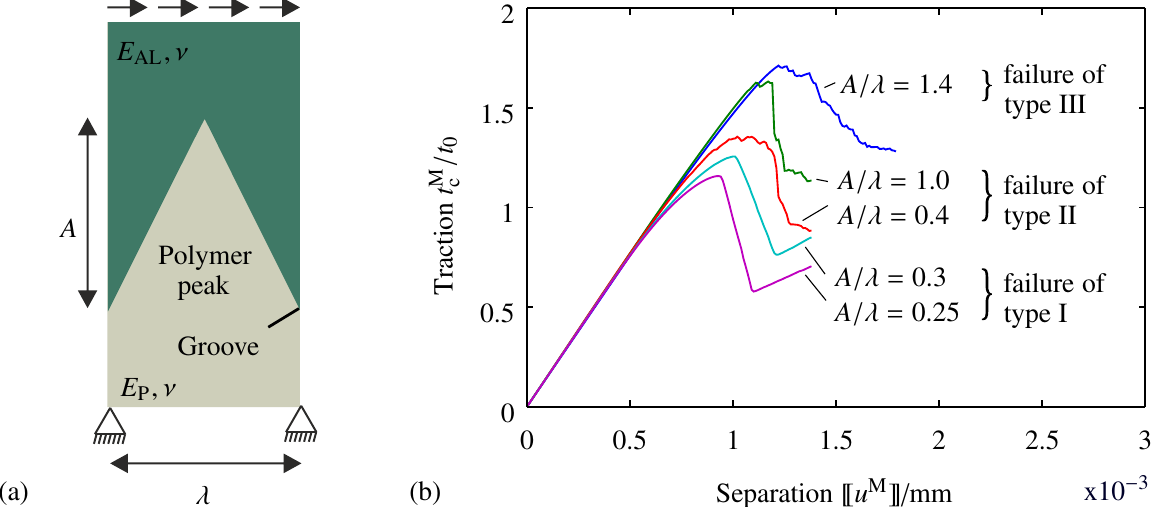}
  \end{footnotesize}
\caption{Macroscopic mode II loading of triangular interface: a) schematic RVE with default configuration and tested loading conditions and b) the homogenized TSL for different roughness ratios $A/\lambda$.}
\label{fig:rve_shear_B}
\end{figure}

\reff{fig:rve_shear_B}~b) shows selected effective shear traction-separation curves. It can be observed that the initial linear slope is followed by a nonlinear hardening regime driven by the initiation and propagation of interface cracks. The maximum tractions and post critical regimes are govern by cohesive failure of the polymer phase. With increasing roughness ratios $A/\lambda$, an extended elastic regime, increased maximum tractions and a tighter nonlinear transition zone from the elastic response to the maximum tractions are found. This embrittlement is caused by the interlocking feature of the interface and was not observed under tension loading in \reff{fig:rve_tension_B}~b). 

\begin{figure}
  \centering
    \begin{footnotesize}
    \includegraphics{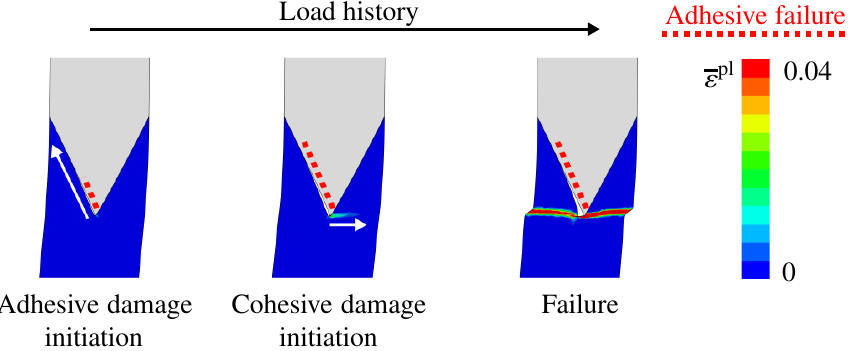}
    \end{footnotesize}
\caption{Macroscopic mode II loading of triangular interface: Evolution of adhesive and cohesive interface failure during load history.}
\label{fig:shear_dmg_B}
\end{figure}

The evolution of adhesive damage and the equivalent plastic strain $\overline\eps^\tx{pl}$ within the default configuration over  the deformation history is shown in \reff{fig:shear_dmg_B}. Similar to mode I loading, adhesive damage initiates due to stress concentrations at the groove of the soft polymer material. The evolving crack results in debonding of the tension side and contact stress transmission at the compression side of the aluminum peak. The equivalent plastic strain localizes due to damage evolution and represents the zones of cohesive failure with $\overline\eps^\tx{pl} \geq 0.04$. Cohesive damage initiates at the groove and localizes horizontally representing a macroscopic crack from one polymer groove to the next one. 

\subsubsection{Influence of roughness enhancement}

\begin{figure}
  \centering
  \begin{footnotesize}
    \includegraphics{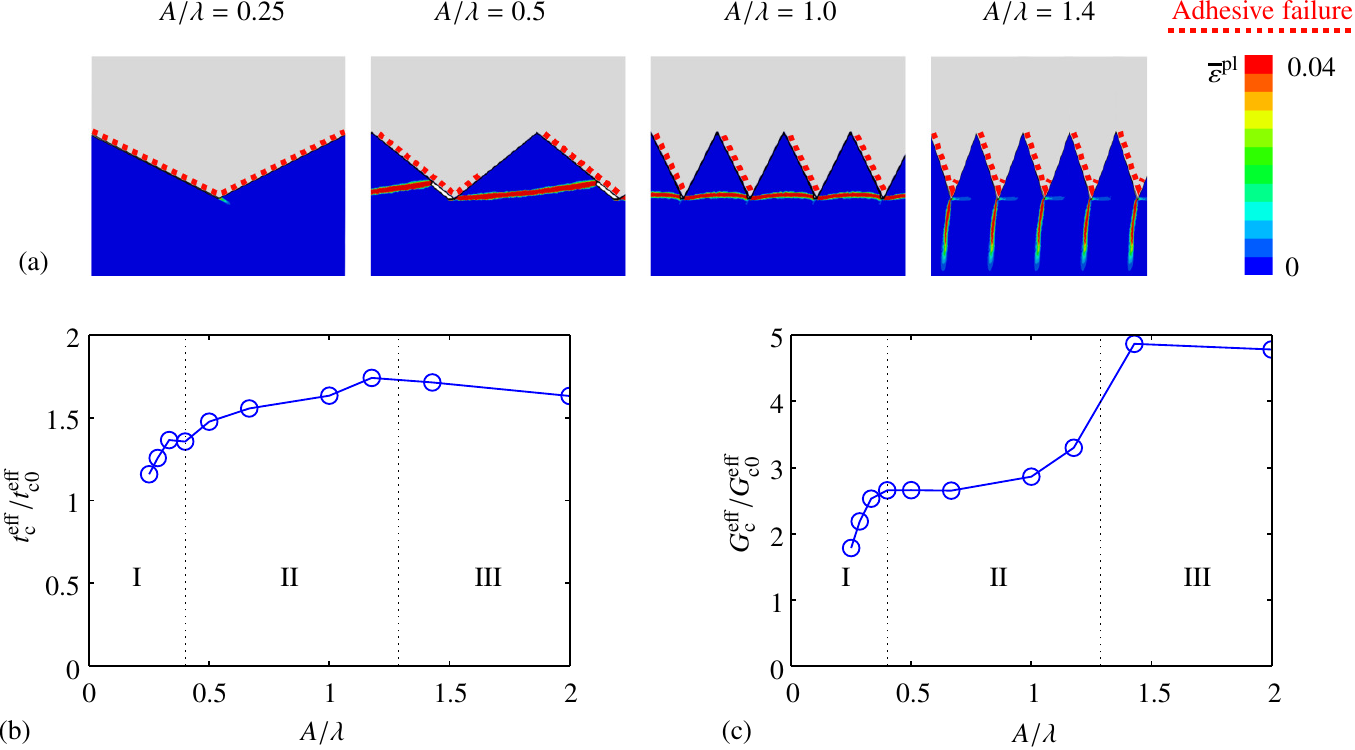}
  \end{footnotesize}
\caption{Macroscopic mode II loading different interface contours:: a) contour plot of localized plastic strains due to damage evolution and b) effective strengths and c) effective critical energies.}
\label{fig:strength_shear_B}
\end{figure}

The shapes of the effective TSLs in \reff{fig:rve_shear_B}~b) obviously depend on the roughness ratio and suggest the definition of three failure types which characterize the interaction of local adhesive and cohesive failure. In order to support these hypotheses, a selection of RVE plots for various roughness ratios are shown in \reff{fig:strength_shear_B}~a). The contour plots present the equivalent plastic strain $\overline\eps^\tx{pl}$ and localized cohesive damage regions $\overline\eps^\tx{pl}>0.4$. For a better visualization, the interface parts with fully developed adhesive damage are marked with a dashed red line.  

\begin{itemize}
  \item \emph{Failure type I}: For low roughness ratios $A/\lambda<1/2$, the pre- and post-critical responses are determined by purely adhesive failure without initiation of cohesive damage. After reaching the maximum traction, the TSL softens due to debonding of the local interface. At the critical separation, e.g. $\se{u}=1.2 \cdot 10^{-3}~\tx{mm}$ for $A/\lambda=1/3$, the complete interface has failed adhesively but the interlock/contact conditions result in an elastic increase of the effective tractions. This sliding process does not cause a full development of cohesive damage. 
\item \emph{Failure type II}: Configurations with intermediate ratios ($1/2\le A/\lambda\le 1.2$) show the expected damage distribution as described in Section \ref{ssec:shear}. Within this roughness range we observe at low ratios, that the cohesive damage path runs from the polymer groove to the opposite peak edge. The damage pattern forms a single macroscopic crack composed of adhesive and cohesive parts. With increasing ratio $A/\lambda$ a shifts from adhesive to fully cohesive failure occurs, indicated by a single cohesive damage path extending from one polymer groove to the next. This failure is called failure of type II.
\item \emph{Failure type III}: For interface configurations with very high roughness ratios ($A/\lambda>1.2$) the exemplary TSL in \reff{fig:rve_shear_B}~b) ($A/\lambda=1.4$) shows a reduced, slowly evolving stress drop over the investigated separation range. This behavior is attributable to the change of localization of cohesive damage which nows initiates at the polymer groove and grows vertically to the bottom of the RVE.
\end{itemize}

For tensile loadings (mode I) it was reasonable to evaluate the effective maximum traction as interface strength and the area under the complete effective TSL (full failure with zero tractions) as fracture energy per unit of created traction free interfaces or critical energy release rate. For mode II loading it is much more difficult to define unique criteria to characterize the effective properties of the interface over large ranges of $A/\lambda$. Unlike the strength evaluation, which can be directly adopted, the definition of failure is not obvious, because it is difficult to load the RVE to the state of zero effective tractions. 

In this study a critical energy is defined to evaluate and compare all failure types, i.e. the area under the curve is assessed up to the point, where the TSL falls below $75\%$ of its maximum traction. The specific value of 75\% ensures to evaluate the critical energy within only the first softening regime for all
considered RVEs. This critical energy could be interpreted differently for the various failure types and can not always be related to the critical energy release rate, but it should allow for a comparison of the softening regime between the different failure types.

The evaluation of the effective interface strength \refe{eq:eff_strength} and critical energy \refe{eq:eff_energy} are plotted versus the corresponding roughness ratio $A/\lambda$ in \reff{fig:strength_shear_B}~b) and c), respectively. Both curves have been normalized to the corresponding properties of the used CZM. At low ratios $A/\lambda$ (failure type I) both quantities increase sharply with a raise in the roughness ratio indicating an enhancement of the bi-material connection similar to the observations under mode I loading. In contrast to pure mode I loading, a significant increase of effective strength can be observed starting from the smallest investigated ratio $A/\lambda=0.25$ which is due to the mechanical interlocking. A further increase of $A/\lambda$ results in failure type II, where the strength increases monotonically with a reduced slope.  At the transition from failure type II to type III the strength reaches a maximum representing an upper bound because the effective strength is determined by the strength of the bulk material. The critical energy for failure of type II increases much slower than for type I. In this range the interpretation of the evaluated critical energy as effective fracture energy per unit crack is possible based on the arguments discussed above. At very high ratios $A/\lambda$ failure of type III occurs, where the effective strength and critical energy show opposite effects. 

\begin{figure}
  \centering
  \begin{footnotesize}
    \includegraphics{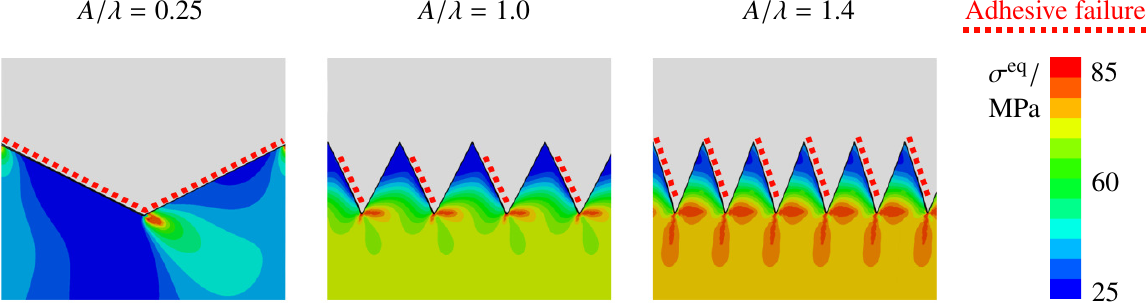}
  \end{footnotesize}
\caption{Contour plot of equivalent von \txsc{Mises} stress and adhesive damage state at different ratios $A/\lambda$ under mode II loading.}
\label{fig:special_B}
\end{figure}

Two main reasons for the occurrence of certain failure types can be identified from the analysis. The first one is the state of previously induced adhesive damage at the point of initiation of cohesive damage and the second one are the induced stress concentrations at the sharp interface corners. Therefore, contour plots of sample RVEs for each failure type are plotted in \reff{fig:special_B}. The von \txsc{Mises} stress distribution is shown as the scalar driving quantity in the yield condition \refe{eq:yield_condition} at the state of cohesive damage initiation. Since the macroscopic mode II loading conditions induce a shear loading, the shear stresses $\sig_\tx{xy}$ dominate the von \tx{Mises} equivalent stress. To illustrate the state of adhesive damage, the edges, where the damage value of the CZM reached nearly one, have been marked with a dashed red line. Failure of type I is characterized by a fully damaged interface, where no shear tractions are transmitted. Hence, the shear stress and the von \txsc{Mises} stress increase only in a very small domain around the polymer groove. As can be seen at $A/\lambda=1$, the general level of $\sig^\tx{eq}$ is much higher compared to $A/\lambda=0.25$. The concentration of shear stresses lies within the polymer peak and causes the peak shear off.  At high roughness ratios a second concentration domain of $\sig^\tx{eq}$, below the polymer groove as shown at $A/\lambda=1.4$ in \reff{fig:special_B}, becomes dominant. This stress concentration is caused by the sharp corner and leads to the changed damage localization of failure type III.

\begin{figure}
  \centering
  \begin{footnotesize}
    \includegraphics{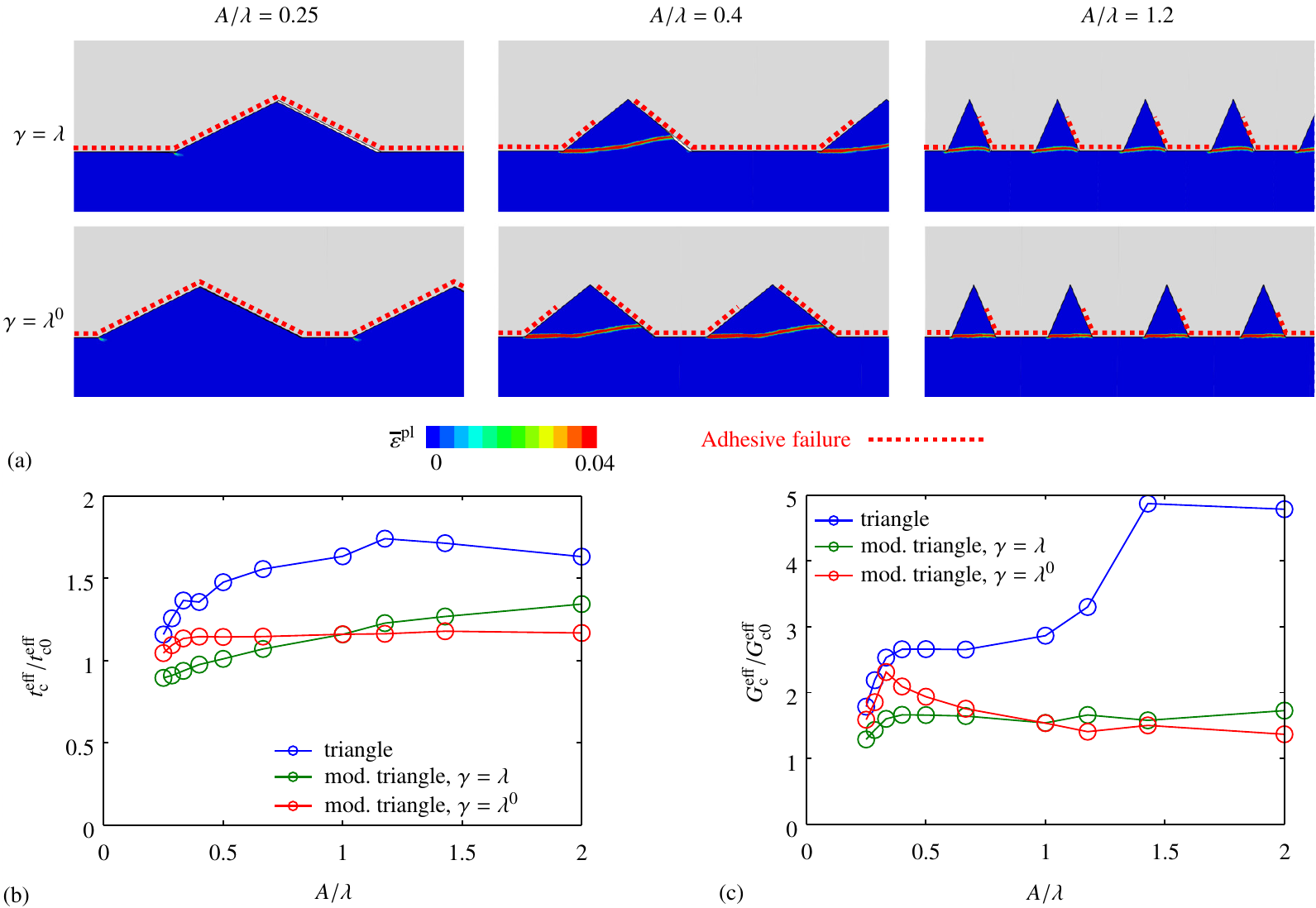}
  \end{footnotesize}
\caption{Macroscopic mode II loading of modified triangular interface: a) contour plot of localized plastic strains due to damage evolution and b) effective strengths and c) effective critical energies.}
\label{fig:strength_tension_A_B}
\end{figure}

\subsubsection{Influence of interface contour}

The final study shows the comparison of the triangular interface with the modified triangular interfaces according to \reff{fig:ideal_profiles}~c). Contour plots of  the localized equivalent plastic strain are shown in \reff{fig:strength_tension_A_B}~a). Similar to the triangular interface, at low ratios $A/\lambda$, failure type I occurs. This process is governed by a fully damaged interface and a sliding contact state without significant cohesive damage evolution in the investigated separation range. With increasing $A/\lambda$ the adhesive damaging process is not yet completed, when cohesive damage initiates. In combination with large edge angles, the characteristic peak shear off of failure type II is observed. In the range of $0.15\leq A/\lambda \leq 2.0$ only failure of type I and type II occurs because the angle between the flat and triangular edges never reaches the critical values of the corresponding triangular shapes. 

\begin{figure}
  \centering
  \begin{footnotesize}
    \includegraphics{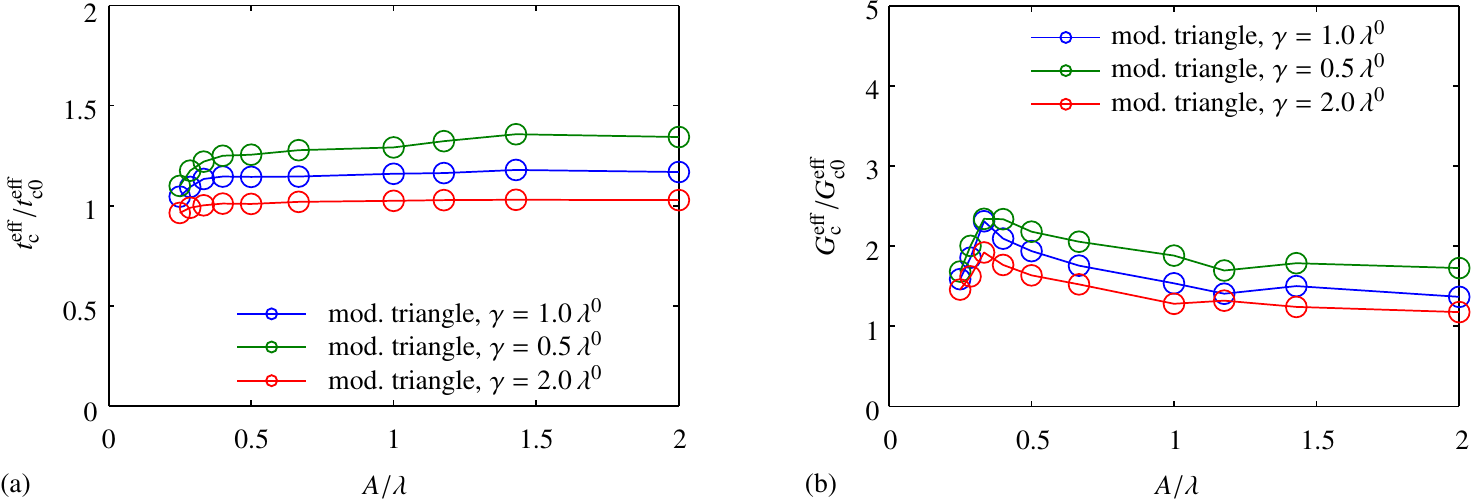}
  \end{footnotesize}
\caption{Influence of geometry parameter $\gamma$ of modified triangular interface under macroscopic mode II loading: a) effective strengths and b) effective critical energies.}
\label{fig:strength_shear_C_mod}
\end{figure}

The effective strength and effective critical energies are plotted in \reff{fig:strength_tension_A_B}~b). The effective strength resembles the characteristics of the triangular interface. An increased roughness $A/\lambda$ has a beneficial effect and leads to an enhanced interface strength.  However, the strength level of the modified triangular interfaces is far below the strength of the triangular interface and at low ratios $A/\lambda<0.4$ the strength even falls below the initial strength of the local CZM. The main reasons are on the one hand, that sharp corners and stress singularities in general reduce the effective strength. On the other hand the flat interface part fails always adhesively due to the weak interface, which attenuates the enhancing effect of the interface roughness. Similar to mode I, the effective strength of the modified interface $\gamma=\lambda^0$ is higher than the case with scaling $\gamma=\lambda$ and can be explained following the discussion in Section \ref{sec:tension_influence_contour}. The effective critical energy raises initially with higher roughness within type I failure for both modified interfaces. In the case of constant $\gamma=\lambda^0$, the flat interface part is small compared to the width of the peak. Hence, a similar increase in critical energy can be observed as the triangular interface. A further increase of $A/\lambda$ leads to failure of type II and constant critical energies for case of scaling $\gamma=\lambda$ and decreasing energies for the modified interface with constant $\gamma=\lambda^0$. The decrease in energy in case of constant $\gamma$ can be explained as follows: Beginning at small ratios $A/\lambda$ (nearly flat), the mechanical interlock dominates the TSL in failure mode I, leading to a quick raise of the critical energy (\reff{fig:strength_tension_A_B}, $\gamma=\lambda^0$, $A/\lambda=0.25$). With the shift to failure type II, the ratio $A/\lambda$ is small but the peak width $\lambda$ is larger than the flat parts $\gamma$, leading to an expanded domain of cohesive failure (\reff{fig:strength_tension_A_B}, $\gamma=\lambda^0$, $A/\lambda=0.4$). A further increase of $A/\lambda$ results in an increase of the flat parts, which fail always adhesively, with respect to a unit interface section (\reff{fig:strength_tension_A_B}, $\gamma=\lambda^0$, $A/\lambda=1.2$). A general enhancement of critical energies at high ratios $A/\lambda$ stays, out due to the missing of failure type III. The influence of parameter $\lambda^0$ is shown in \reff{fig:strength_shear_C_mod}. For various $\lambda^0$ the effective strength and critical energy are shifted on the ordinate. Higher values of $\lambda^0$ result in an increase of the flat area with  adhesive failure and thus in decreased effective interface properties.

\section{Conclusion}

In this work a modeling strategy was presented to study the failure behavior of bi-material interfaces, whose shape may be contoured representing a connection of rough or structured surfaces. The presented method enables the modeling of interface failure and the extraction of effective traction-separation relations of bi-material interfaces. Case studies show the influence of individual roughness parameters of the bi-material interface and demonstrate the general capability of the modeling strategy. The simulations indicate that it is possible to adjust the effective adhesion properties by increasing the interface roughness. The main mechanism leading to a strengthening of the joint is the transition from adhesive to cohesive failure type. The presented study was performed with a generic configuration representing an aluminum interface connected to a much more softer polymer material. The objective of constitutive model selection and parametrization was to describe typical bi-material characteristics with widely used models. The general occurrence and ratios of adhesive and cohesive failure depends strongly on the specific material combination. Hence, model extensions to capture viscoplastic effects or the large deformation of the polymer are conceivable.\\[24pt]

{\bf Acknowledgments:}

The present project is supported by the German Research Foundation (DFG) within the Priority Program (SPP) 1712, KA3309/4-1. This support is gratefully acknowledged. We thank the Institute of Lightweight Construction and Polymer Technology (ILK) at TU Dresden for providing surface measurement data of the rough interface.

\bibliography{references}

\begin{thebibliography}{10}
\expandafter\ifx\csname url\endcsname\relax
  \def\url#1{\texttt{#1}}\fi
\expandafter\ifx\csname urlprefix\endcsname\relax\def\urlprefix{URL }\fi
\expandafter\ifx\csname href\endcsname\relax
  \def\href#1#2{#2} \def\path#1{#1}\fi

\bibitem{ashby2003}
M.~Ashby, Y.~Brechet, Designing hybrid materials, Acta materialia 51~(19)
  (2003) 5801--5821.

\bibitem{grujicic2008}
M.~Grujicic, V.~Sellappan, M.~Omar, N.~Seyr, A.~Obieglo, M.~Erdmann,
  J.~Holzleitner, An overview of the polymer-to-metal direct-adhesion hybrid
  technologies for load-bearing automotive components, Journal of Materials
  Processing Technology 197~(1) (2008) 363--373.

\bibitem{yao2002}
Q.~Yao, J.~Qu, Interfacial versus cohesive failure on polymer-metal interfaces
  in electronic packaging—effects of interface roughness, Journal of
  Electronic Packaging 124~(2) (2002) 127--134.

\bibitem{zhao2007}
X.-L. Zhao, L.~Zhang, State-of-the-art review on frp strengthened steel
  structures, Engineering Structures 29~(8) (2007) 1808--1823.

\bibitem{bruzzone2008}
A.~Bruzzone, H.~Costa, P.~Lonardo, D.~Lucca, Advances in engineered surfaces
  for functional performance, CIRP Annals-Manufacturing Technology 57~(2)
  (2008) 750--769.

\bibitem{lucchetta2011}
G.~Lucchetta, F.~Marinello, P.~Bariani, Aluminum sheet surface roughness
  correlation with adhesion in polymer metal hybrid overmolding, CIRP
  Annals-Manufacturing Technology 60~(1) (2011) 559--562.

\bibitem{kim2010}
W.-S. Kim, I.-H. Yun, J.-J. Lee, H.-T. Jung, Evaluation of mechanical interlock
  effect on adhesion strength of polymer--metal interfaces using
  micro-patterned surface topography, International Journal of Adhesion and
  Adhesives 30~(6) (2010) 408--417.

\bibitem{Cordisco2016}
F.~A. Cordisco, P.~D. Zavattieri, L.~G. Hector~Jr., B.~E. Carlson, Mode i
  fracture along adhesively bonded sinusoidal interfaces, International Journal
  of Solids and Structures (2016) --.

\bibitem{noijen2009}
S.~Noijen, O.~van~der Sluis, P.~Timmermans, G.~Zhang, Numerical prediction of
  failure paths at a roughened metal/polymer interface, Microelectronics
  Reliability 49~(9) (2009) 1315--1318.

\bibitem{Zavattieri2007}
P.~D. Zavattieri, L.~G. Hector, A.~F. Bower, Determination of the effective
  mode-i toughness of a sinusoidal interface between two elastic solids,
  International Journal of Fracture 145~(3) (2007) 167--180.

\bibitem{Kaestner2016}
M.~K\"astner, S.~M\"uller, F.~Hirsch, J.-S. Pap, I.~Jansen, V.~Ulbricht, {XFEM}
  modeling of interface failure in adhesively bonded fiber-reinforced polymers,
  Advanced Engineering Materials 18~(3) (2016) 417--426.

\bibitem{Zavattieri2008}
P.~D. Zavattieri, L.~G. Hector~Jr., A.~F. Bower, Cohesive zone simulations of
  crack growth along a rough interface between two elastic-plastic solids,
  Engineering Fracture Mechanics 75~(15) (2008) 4309--4332.

\bibitem{Cordisco2012}
F.~A. Cordisco, P.~D. Zavattieri, L.~G. Hector~Jr., A.~F. Bower, Toughness of a
  patterned interface between two elastically dissimilar solids, Engineering
  Fracture Mechanics 96 (2012) 192--208.

\bibitem{Cordisco2014}
F.~Cordisco, P.~D. Zavattieri, L.~G. Hector~Jr., A.~F. Bower, On the mechanics
  of sinusoidal interfaces between dissimilar elastic-plastic solids subject to
  dominant mode i, Engineering Fracture Mechanics 131 (2014) 38--57.

\bibitem{Matous2008}
K.~Matou\v{s}, M.~G. Kulkarni, P.~H. Geubelle, Multiscale cohesive failure
  modeling of heterogeneous adhesives, Journal of the Mechanics and Physics of
  Solids 56~(4) (2008) 1511--1533.

\bibitem{Alfaro2010}
M.~Cid~Alfaro, A.~Suiker, C.~Verhoosel, R.~de~Borst, Numerical homogenization
  of cracking processes in thin fibre-epoxy layers, European Journal of
  Mechanics - A/Solids 29~(2) (2010) 119--131.

\bibitem{Verhoosel2010}
C.~V. Verhoosel, J.~J.~C. Remmers, M.~A. Gutiérrez, R.~de~Borst, Computational
  homogenization for adhesive and cohesive failure in quasi-brittle solids,
  Int. J. Numer. Meth. Engng. 83~(8-9) (2010) 1155--1179.

\bibitem{Palmieri2014}
V.~Palmieri, L.~De~Lorenzis, Multiscale modeling of concrete and of the
  {FRP}-concrete interface, Engineering Fracture Mechanics 131 (2014) 150--175.

\bibitem{hirschberger2009}
C.~Hirschberger, S.~Ricker, P.~Steinmann, N.~Sukumar, Computational multiscale
  modelling of heterogeneous material layers, Engineering Fracture Mechanics
  76~(6) (2009) 793--812.

\bibitem{hill1967}
R.~Hill, The essential structure of constitutive laws for metal composites and
  polycrystals, Journal of the Mechanics and Physics of Solids 15~(2) (1967)
  79--95.

\bibitem{vossen2014}
B.~G. Vossen, P.~J. Schreurs, O.~van~der Sluis, M.~Geers, Multi-scale modeling
  of delamination through fibrillation, Journal of the Mechanics and Physics of
  Solids 66 (2014) 117--132.

\bibitem{Kachanov1985}
L.~M. Kachanov, Time of the rupture process under creep conditions., Izv. Akad.
  Nauk. S.S.R. Otd. Tech. Nauk 8.

\bibitem{Hillerborg1976}
A.~Hillerborg, M.~Modéer, P.-E. Petersson, Analysis of crack formation and
  crack growth in concrete by means of fracture mechanics and finite elements,
  Cement and Concrete Research 6~(6) (1976) 773 -- 781.

\bibitem{Valoroso2006}
N.~Valoroso, L.~Champaney, A damage-mechanics-based approach for modelling
  decohesion in adhesively bonded assemblies, Engineering Fracture Mechanics
  73~(18) (2006) 2774--2801.

\bibitem{Camanho2003}
P.~P. Camanho, C.~G. Davila, M.~F. de~Moura, Numerical simulation of mixed-mode
  progressive delamination in composite materials, Journal of Composite
  Materials 37~(16) (2003) 1415--1438.

\bibitem{Temizer2011}
I.~Temizer, Thermomechanical contact homogenization with random rough surfaces
  and microscopic contact resistance, Tribology International 44~(2) (2011)
  114--124.

\end{thebibliography}

\end{document}